\documentclass[usenatbib]{mn2e}

\bibpunct{(}{)}{;}{a}{}{,}

\newcommand{\ergcm}[1]{$\times 10^{#1}$ erg cm$^{-2}$ s$^{-1}$}

\newcommand{\ergs}[1]{$\times 10^{#1}$ erg s$^{-1}$}

\newcommand{\hcm}[1]{$\times 10^{#1}$ cm$^{-2}$}

\newcommand{\ct}{cts s$^{-1}$}

\newcommand{\ltsima}{$\buildrel < \over \sim$}
\newcommand{\lsim}{\lower.5ex\hbox{\ltsima}}
\newcommand{\gtsima}{$\buildrel > \over \sim$}
\newcommand{\gsim}{\lower.5ex\hbox{\gtsima}}

\newcommand{\xmmp}{\hbox{XMMU\,J005011.2-730026}}

\usepackage{graphicx}
\usepackage{amssymb}
\usepackage{epsfig}
\usepackage{natbib}
\def\ion#1#2{#1$\;${\small\rm\@Roman{#2}}\relax}

\title[SXP214]{The XMM-Newton survey of the Small Magellanic Cloud:\\
       XMMU\,J005011.2-730026 = SXP214, a Be/X-ray binary pulsar
       \thanks{Based on observations with
               XMM-Newton, an ESA Science Mission with instruments and contributions
               directly funded by ESA Member states and the USA (NASA)}}

\author[M.J. Coe et al.]{M. J.~Coe$^{1}$, F. Haberl$^{2}$, R. Sturm$^{2}$, W.~Pietsch$^{2}$, L.J. Townsend$^{1}$, E.S. Bartlett$^{1}$,\and M. Filipovic$^{3}$, A. Udalski$^{4}$, R.H.D. ~Corbet$^{5}$, A. Tiengo$^{6}$, M. Ehle$^{7}$, J.L. Payne$^{3}$ \and \& D. Burton$^{8}$  \\
$^{1}$ School of Physics and Astronomy, University of Southampton, SO17
1BJ, UK\\
$^{2}$ Max-Planck-Institut f\"ur extraterrestrische Physik,
           Giessenbachstra{\ss}e, 85748 Garching, Germany\\
$^{3}$ University of Western Sydney, Locked Bag 1797, Penrith South DC, NSW1797, Australia \\
$^{4}$ Warsaw University Observatory, Aleje Ujazdowskie 4, 00-478 Warsaw, Poland \\
$^{5}$ University of Maryland, Baltimore County, Mail Code 662, NASA Goddard Space Flight Center, Greenbelt, MD 20771, USA  \\
$^{6}$ INAF - Istituto di Astrofisica Spaziale e Fisica Cosmica - Milano, via E. Bassini 15, I-20133 Milano, Italy. \\
$^{7}$ European Space Agency, XMM-Newton Science Operations Centre, P.O. Box 78, 28691 Villanueva de la Canada, Madrid, Spain. \\
$^{8}$ University of Southern Queensland, Toowoomba Qld 4350, Australia. \\
}

\begin{document}

\date{18 February 2011}

\pagerange{\pageref{firstpage}--\pageref{lastpage}} \pubyear{2002}

\maketitle

\label{firstpage}

\begin{abstract}

{   In the course of the XMM-Newton survey of the Small Magellanic Cloud (SMC), a region
   to the east of the emission nebula N19 was observed in November 2009.
   To search for new candidates for high mass X-ray binaries the EPIC PN and MOS data of the detected
   point sources were investigated and their spectral and temporal characteristics identified. A new transient (\xmmp = SXP214) with a pulse period of 214.05 s was discovered; the source had a hard X-ray
   spectrum  with power-law index of $\sim$0.65. The accurate X-ray source location permits the identification of the X-ray source with a $\sim$15th magnitude Be star, thereby confirming this system as a new Be/X-ray binary.

}

\end{abstract}


\section{Introduction and background}

The Be/X-ray systems represent the largest sub-class of all High Mass X-ray Binaries (HMXB).  A survey of the literature reveals that of the $\sim$240 HMXBs known in our Galaxy and the Magellanic Clouds (Liu et al., 2005, 2006), $\ge$50\%
fall within this class of binary.  In fact, in recent years it has emerged that there is a substantial population of HMXBs in the SMC comparable in number to the Galactic population. Though unlike the Galactic population, all except one of the SMC HMXBs are Be star systems.  In these systems the orbit of the Be star
and the compact object, presumably a neutron star, is generally wide
and eccentric.  X-ray outbursts are normally associated with the
passage of the neutron star close to the circumstellar disk (Okazaki
\& Negueruela 2001), and generally are classified as Types I or II (Stella, White \& Rosner, 1986). The Type I outbursts occur periodically at the time of the periastron passage of the neutron star, whereas Type II outbursts are much more extensive and occur when the circumstellar material expands to fill most, or all of the orbit. This paper concerns itself with Type I outbursts. General reviews of such HMXB systems may
be found in Negueruela (1998), Corbet et al. (2008) and Coe et al. (2000, 2008).

One of the aims of the XMM-Newton
large program SMC survey (Haberl \& Pietsch 2008a) is the ongoing
study of the Be/X-ray binary population of the SMC,
which can be used as a star formation tracer for $\sim$50
Myr old populations (Antoniou et al. 2010). In this paper we
present the analysis of X-ray and optical data from the newly
discovered X-ray pulsar \xmmp, hereafter referred to more simply as SXP214 following the naming convention of Coe et al (2005) for X-ray binary pulsars in the SMC.

\section{X-ray Observations}

The local group galaxies are best suited to study their X-ray source populations using present day observatories. Extending the existing archival observations, we have carried out a large program in 2009 with XMM-Newton to obtain a complete X-ray survey of the SMC in the 0.1-10 keV band. The hard X-ray source SXP214 that is the subject of this work was discovered as transient in observation 14 (observation ID 0601211401 of the survey), on 4 Nov. 2009. The field is centred  about 22\arcmin\ east of the emission nebula N19 in the south-western part  of the SMC  bar.
It was observed by all three EPIC instruments (Str{\"u}der et al., 2001, Turner et al., 2001)  at off-axis angles between 12.9\arcmin\ and 13.7\arcmin\
on CCD 10, 3, and 4 for EPIC-pn, EPIC-MOS1, and EPIC-MOS2, respectively.
All three cameras were operated in full-frame imaging mode with CCD
readout frame times of 73 ms (pn)
and 2.6 s (MOS).
We used the XMM-Newton Science Analysis System (SAS) version 10.0.0\footnote{http://xmm.esac.esa.int/sas/} to reduce the data.

To correct the astrometric bore-sight, we identified eight sources (mainly known Be/X-ray binaries) in the field of view (FoV) with the Magellanic Clouds Photometric
Survey of Zaritsky et al. (2002) and obtained a shift of $\Delta$RA=0.10\arcsec\ and $\Delta$Dec=0.70\arcsec.
The corrected X-ray position as found by {\tt emldetect} is
R.A. = 00$^{\rm h}$50$^{\rm m}$10\fs95 and Dec. = --73\degr00\arcmin25\farcs0 (J2000.0),
with a statistical error of 0.18\arcsec\ and a systematic uncertainty of $\sim$1\arcsec\ (1 $\sigma$ confidence for both cases). The relatively large systematic error is caused by the large off-axis angle of the source.

To remove times affected by background flares due to soft protons which occured at the end of the observation (performed at the end of the satellite orbit)
we defined good time intervals by applying thresholds on the background count rate in the 7.0--15.0 keV band of 8 and 2.5 cts ks$^{-1}$ arcmin$^{-2}$
for EPIC-pn and EPIC-MOS, respectively. The soft proton background was at a very low level during the first part of the observation,
thus resulting in net exposure times of 31.4 ks and 32.7 ks for EPIC-pn and EPIC-MOS, respectively.

\begin{figure}
  \resizebox{\hsize}{!}{\includegraphics[angle=0,clip=]{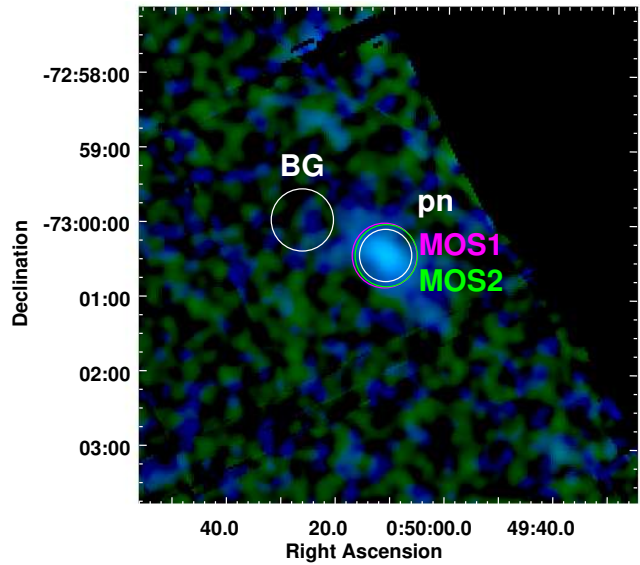}}
  \caption{EPIC colour image of SXP214 combining pn and MOS data. The red, green and blue colours represent the X-ray intensities in the
           0.2$-$1.0, 1.0$-$2.0 and 2.0$-$4.5 keV energy bands. Circles indicate the extraction regions
	   (with radii of 21\arcsec, 26\arcsec, and 25\arcsec\ for pn, MOS1, and MOS2 source regions and 25\arcsec\ for the background). }
  \label{fig:ima}
\end{figure}

To define extraction regions (see Fig.~\ref{fig:ima}) for the source and background with optimised signal to noise ratio, the SAS task {\tt eregionanalyse} was used.
We ensured that the source extraction region had a distance of $>$10\arcsec\ to other detected sources.
For the background extraction, we defined a circular region covering the same point source free area on the sky for all three detectors. To avoid systematic detector background variations present close to CCD borders where the source was located, we chose an area  on the same CCDs as the source, and  in the case of pn  at a similar distance to the readout node. This restricted our selection to a radius of 25\arcsec.
We extracted spectra, by selecting single and double pixel events from the EPIC-pn data and single to quadruple
events from EPIC-MOS data, both with {\tt FLAG = 0}.
The EPIC-pn/MOS1/MOS2 spectra contain 989/464/396 background subtracted counts in the 0.2--10.0 keV band, respectively, and were binned to a minimum signal-to-noise ratio of 5 for each bin. To increase the statistics for the timing analysis we also generated a merged event list from all three instruments, containing 2053 cts (source + background).

\subsection{Spectral analysis}

\begin{figure}
  \resizebox{\hsize}{!}{\includegraphics[angle=-90,clip=]{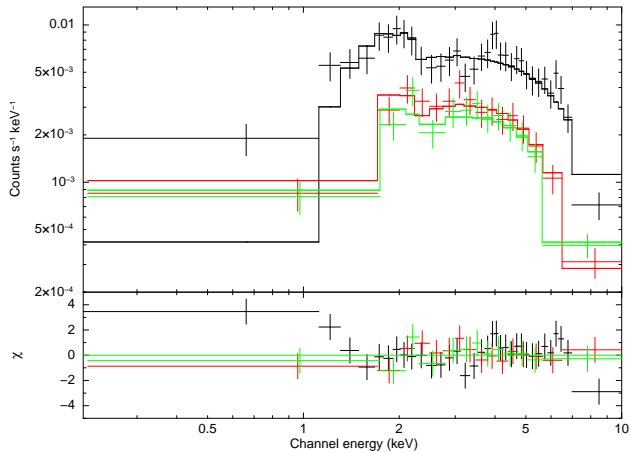}}
  \caption{EPIC spectra of SXP214. The top panel shows the EPIC-pn (black), EPIC-MOS1 (red), and EPIC-MOS2 (green) spectra together with the best-fit model.
           The residuals are shown in the lower panel.
          }
  \label{fig:spec}
\end{figure}

We used {\tt XSPEC} (Arnaud et al, 1996) version 12.6.0k for spectral fitting.
The three EPIC spectra (see Fig.~\ref{fig:spec}) were fitted simultaneously with a common set of spectral model parameters
and a relative normalisation factor for instrumental differences.
The Galactic photo-electric foreground absorption  was set to a column density  of N$_{\rm H{\rm , GAL}}$ = 6\hcm{20}
with abundances according to Wilms et al (2000).
The SMC column density was a free parameter with abundances for elements heavier than Helium fixed at 0.2 (Russell \& Dopita 1992).
The emission was modelled with an absorbed power-law where we obtained
an SMC column density N$_{\rm H, SMC} = (4.96\pm1.50) \times 10^{22}$ cm$^{-2}$, a photon-index $\Gamma = 0.65\pm0.18$ and a
flux in the 0.2--10.0 keV band of $(1.2\pm0.2)\times 10^{-12}$ erg cm$^{-2}$ s$^{-1}$.
Assuming a distance of 60 kpc, this corresponds to an unabsorbed luminosity of $7.1\times 10^{35}$ erg s$^{-1}$.
The model resulted in an acceptable fit with $\chi^2/{\rm dof} = 57/58$, with relative normalisation factors
c$_{\rm MOS1} =1.18\pm0.13$, c$_{\rm MOS2}=1.03\pm0.12 $ (relative to c$_{\rm pn} = 1$).

Since the EPIC-pn spectrum indicates a weak excess at $\sim$6.5 keV, we investigated a possible contribution of iron K emission lines at 6.4 keV (fluorescent emission) and 6.7 keV (Fe XXV)  by assuming an unresolved line width.
We obtained 90\% upper limits of 4.0$\times 10^{-6}$ photons cm$^{-2}$ s$^{-1}$ and 3.0$\times 10^{-6}$ photons cm$^{-2}$ s$^{-1}$, which
correspond to equivalent width upper limits of 215 eV and 163 eV, respectively.

\subsection{Timing analysis}

\begin{figure}
  \resizebox{\hsize}{!}{\includegraphics[angle=-90,clip=]{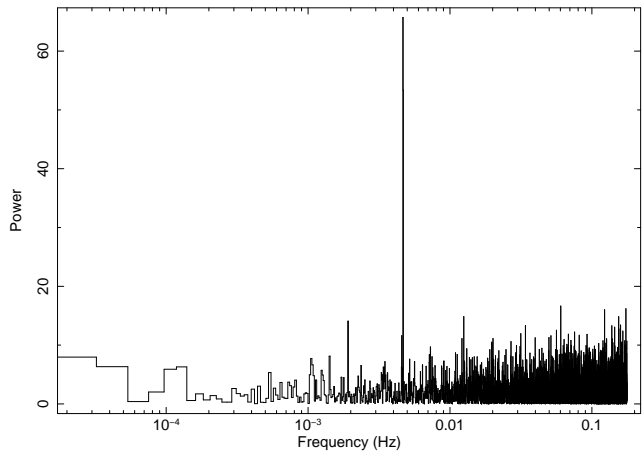}}
  \caption{Power density spectrum created from the merged EPIC data in the 0.2--10.0 keV energy band.
           The time binning of the input light curve is 2.840 s. }
  \label{fig:psd}
\end{figure}

\begin{figure}
  \begin{center}
  \resizebox{0.9\hsize}{!}{\includegraphics[angle=0,clip=]{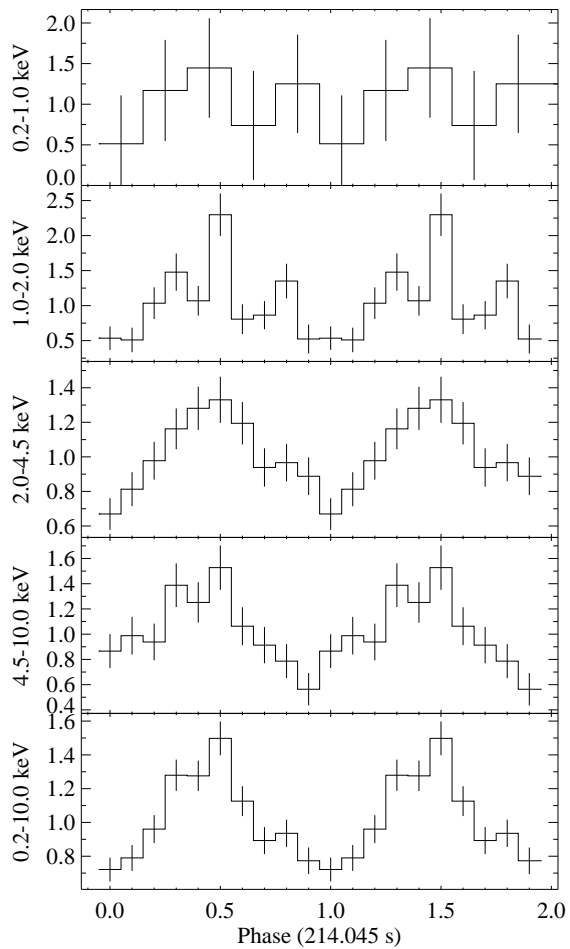}
            }
  \end{center}
  \caption{Pulse profiles obtained from the merged EPIC-pn/MOS1/MOS2 data in different energy bands
           (for better statistics the first two standard energy bands were combined in the top panel,
            the bottom panel shows all five energy bands combined).
	   The profiles are background-subtracted and normalized to the
           average count rate (0.0013, 0.0086, 0.027, 0.017 and 0.056 \ct, from top to bottom.
          }
  \label{fig:pp}
\end{figure}

The photon arrival times were randomised within the detector CCD frame time and recalculated for the solar system barycentre using the SAS task {\tt barycen}.
We searched for pulsations in the X-ray light curves in the EPIC standard energy bands (0.2-0.5 keV, 0.5-1.0 keV, 1.0-2.0 keV, 2.0-4.5 keV and
4.5-10 keV) and combinations of them, using fast Fourier transform (FFT) and light curve folding techniques.
The power density spectra derived from light curves in various energy bands from the different EPIC instruments showed a periodic signal at 0.00467~Hz.
To maximise the signal to noise ratio, we created light curves from the merged EPIC event list (filtered by the GTIs common to EPIC-pn and -MOS).
Figure~\ref{fig:psd} shows the inferred power density spectrum from the 0.2-10.0 keV energy band with the clear peak at a frequency of 0.00467~Hz.
Following Haberl, Eger \& Pietsch (2008) we used a Bayesian periodic signal detection method  (Gregory \& Loredo, 1996) to determine the pulse period
with 1$\sigma$ error to $(214.045\pm 0.052)$ s.
In Fig.~\ref{fig:pp}, pulse profiles folded with this period in different energy bands
(based on the EPIC standard bands)
are shown.
Clear variations are seen above 1 keV, while due to the high absorption at lower energies the count rate is insufficient to detect a significant modulation.
Hardness ratios were derived from the pulse profiles in two adjacent energy bands
(HR$_{i}$ = (R$_{i+1}$ $-$ R$_{i}$)/(R$_{i+1}$ + R$_{i}$) with R$_{i}$ denoting the background-subtracted
count rate in energy band $i$ (with $i$ from 1 to 4). Given the relatively low count rate, no significant hardness ratio variations are seen.
We determined a pulsed fraction of $(29 \pm 9)$\% for the 0.2$-$10.0 keV band, assuming a sinusoidal pulse profile.

\subsection{Long-term X-ray variability}

The position of SXP214 was observed with XMM-Newton three times before our large programme SMC survey,
without any detection of the source.
To derive upper limits, we used sensitivity maps of EPIC-pn
and derived 3$\sigma$ upper limits in the 0.2--10.0 keV band of
5.3$\times 10^{-3}$ \ct\ (ObsID: 0110000101, 2000 Oct. 15),
1.1$\times 10^{-2}$ \ct\ (ObsID: 0404680101, 2006 Oct. 5),  and
4.1$\times 10^{-3}$ \ct\ (ObsID: 0403970301, 2007 Mar. 12).
Assuming the best fit power-law spectrum from above, this corresponds to absorbed flux limits of
5.0\ergcm{-14},
1.0\ergcm{-13}, and
3.9\ergcm{-14}
or absorption corrected luminosity limits of
2.5\ergs{34},
5.0\ergs{34}, and
1.9\ergs{34}, respectively.
The second limit is higher, due to a shorter background screened exposure of $\sim$7.7 ks,
compared to $\sim$21 ks for the other two observations.
The upper limits show that SXP214 increased in brightness at least by a factor of 30 during its outburst.

\begin{figure}
\includegraphics[angle=-0,width=80mm]{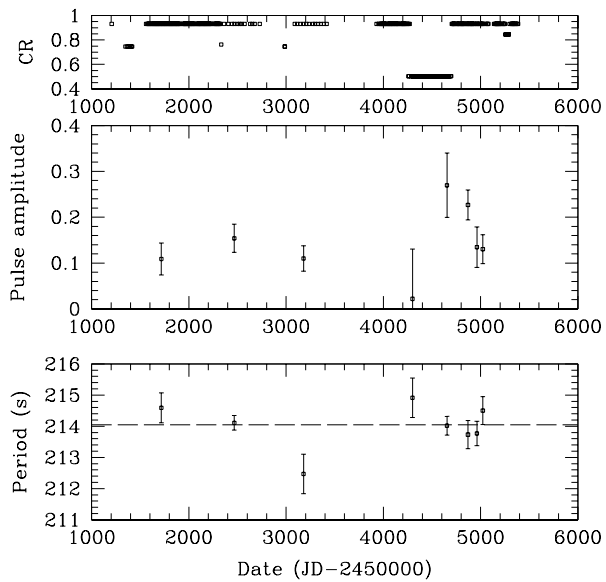}
\caption{RXTE detections of SXP214. The top plot shows the RXTE collimator response (CR) for the periods of time when this value exceeded 0.5. The middle plot shows the pulsed amplitude in counts/PCU/s, and the bottom plot shows the period detected. The dashed line in the bottom plot indicates the XMM-Newton period reported in this work. }
\label{fig:xte}
\end{figure}

SXP214 has frequently fallen within the field of view of RXTE as part of the long-term monitoring of the SMC being carried out by some of the authors on this paper (see Galache et al, 2008 for a discussion of this campaign). Though the collimator sensitivity to SXP214 has been very high for periods of many months at a time, there are only eight possible detections with a confidence $\ge$99\% at a high collimator response of $\ge$0.5 in over 10 years. These detections are shown in Figure~\ref{fig:xte}. Unfortunately, the sparsity of these detections do not give us any significant insight into the binary period of this system. Neither do they reveal any obvious long term spin period change.

The RXTE pulse amplitude may be converted to luminosity assuming a distance of 60kpc to the SMC (though the depth of the sources within the SMC is unknown and \emph{may} affect this distance by up to $\pm$10kpc). The X-ray spectrum was assumed to be a power law with a photon index = 1.5 and an $N_{H}=6\times10^{20}cm^{-2}$. Furthermore it was assumed that there was an average pulse fraction of 33\% for all the measurements and hence the correct total flux is 3 times the pulse component. Thus the luminosity may be determined from the pulse amplitude values shown in Figure~\ref{fig:xte} using the relationship:
\\
\\
$L_X$ = 0.4 $\times\ 10^{37}$ $\times$ 3$R$  erg~s$^{-1}$
\\
\\ where $R$ = pulsed amplitude counts in units of PCU$^{-1}$ s$^{-1}$.

The conclusion from the RXTE data is that there have been no major X-ray outbursts ($L_{x}\ge10^{37}$ erg/s) over the last decade from SXP214.

\section{The optical counterpart}

\subsection{Optical \& IR photometry}

\begin{figure}
\includegraphics[angle=-0,width=80mm]{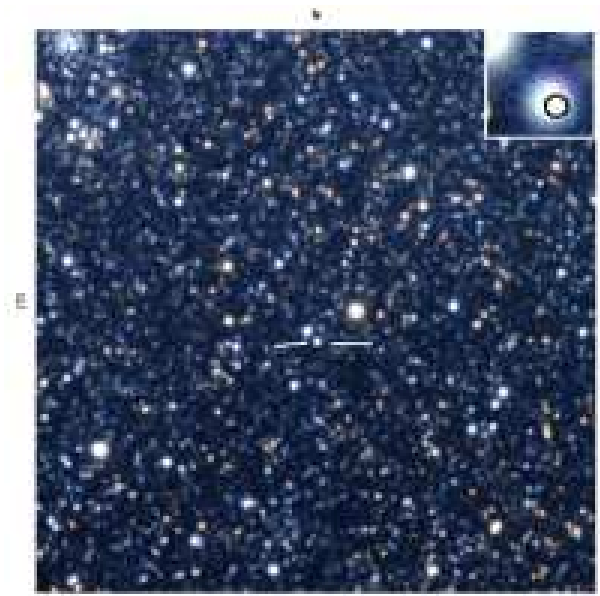}
\caption{Finding chart for SXP214=\xmmp. The colour image size is 5 x 5 arcminutes constructed from the B, V \& R images taken with the Faulkes Telescope. The optical counterpart is indicated. An enlargement of the central 10 x 10 arcsec is shown in the top right corner, with the location and size of the X-ray error circle indicated. }
\label{fig:fc}
\end{figure}

\begin{figure}
\includegraphics[angle=-0,width=80mm]{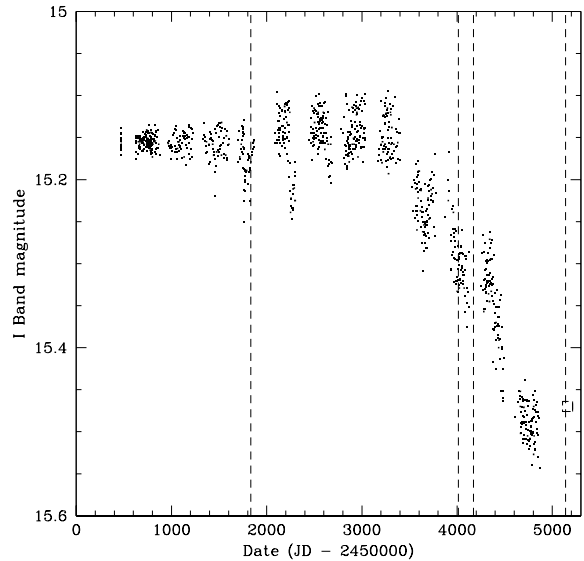}
\caption{Optical I band measurements of the counterpart to SXP214 from OGLE II and III. The last point (open box) comes from the Faulkes Telescope taken on MJD 55160. The dates of the XMM observations are indicated by the vertical dashed lines - the source was not detected in the first three, just in the fourth observation (MJD55139). }
\label{fig:ogle1}
\end{figure}

 From its precise X-ray position the optical counterpart to SXP214 is identified as SMC SC5 207965 in OGLE II and SMC100.3 36998 in OGLE III (Udalski et al, 1997)- see Figure~\ref{fig:fc}. It is also present in the MACHO data as object 212.15966.18, but since those data significantly overlap in time with the OGLE II data, and are not taken through a standard I band filter, they are not discussed any further.  The long term optical measurements of the counterpart to SXP214 are presented in Figure~\ref{fig:ogle1}.

An extra I band point was obtained as part of a set of BVRI observations made using the Faulkes Telescope on 25 Nov 2009 (MJD 55160) - 21d after the XMM detection reported here. The Faulkes Telescope
is located at Siding Spring, Australia and is a 2m, fully autonomous,
robotic Ritchey-Chretien reflector on an alt-azimuth
mount. The telescope employs a Robotic Control System (RCS).
The telescope was used in Real Time Interface mode for the
observation of SXP214.  All the observations
were pipeline-processed (flat-fielding and de-biasing of the images).
The magnitudes of the optical counterpart in all the observed bands were determined
by comparison with several
other nearby stars on the same image frame. In the case of the I-band stars from the OGLE
database were used - these comparison stars have not exhibited any significant
variability in the last 8 years of OGLE monitoring. In all other wavebands reference stars from the catalogue of Massey (2002) were used.

The IR counterpart is identified as Sirius 00501126-7300260 (Kato et al, 2007) and 2MASS 00501125-7300260. The Sirius observations took place on 31 August 2002 and obtained the following results: J = 15.32$\pm$0.02, K = 15.37$\pm$0.02. New measurements in the J, H \& K photometric
bands were obtained on 2009 Dec 11 (MJD 55176) with the same Sirius camera on
the 1.4m IRSF telescope in South Africa (Kato et al. 2007). The full set of photometric measurements are shown in Table~\ref{tab:optir} and reveal that the IR magnitudes were fainter by $\sim$0.3 at the time of the XMM-Newton detection compared with the catalogue numbers.

\begin{table}
\caption{Optical \& IR photometric measurements of SXP214 taken in Nov and Dec 2009 - see text for details.}
\label{tab:optir}
 \begin{tabular}{lll}
  \hline
   Waveband & Magnitude &Error on Magnitude   \\
  \hline
B&15.22&0.02 \\
V&15.33&0.02 \\
R& 15.47 & 0.01 \\
I & 15.47 & 0.03 \\
J & 15.67 & 0.02 \\
H & 15.55 & 0.03 \\
K & 15.61 & 0.09 \\
 \hline
 \end{tabular}
\end{table}

The (B-V) colour index obtained from our data is
(B-V)=--0.11$\pm$0.03. Correcting for an extinction to the SMC of
E(B-V)=0.09 (Schwering \& Israel 1991) gives an intrinsic colour of (B-V)=--
0.20$\pm$0.03. From Wegner (1994) this indicates a spectral type
in the range B1V - B3V - typical of optical counterparts to
Be/X-ray binaries in the SMC (McBride et al. 2008). However,
care must always be taken when interpreting colour information
as a spectral type in systems that clearly have circumstellar disks. Such disks can make significant contributions
to the B and V bands.

\begin{figure}
\includegraphics[angle=-0,width=80mm]{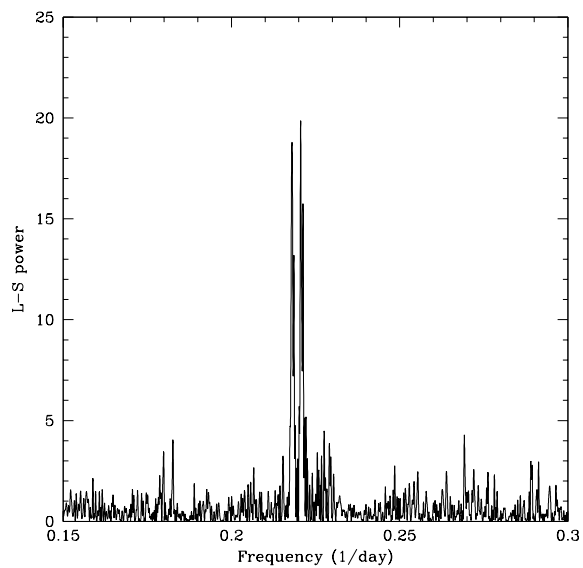}
\caption{Lomb-Scargle power spectrum of the OGLE III data set. }
\label{fig:ogle2}
\end{figure}

The OGLE III data were detrended using a polynomial function and then subjected to a period search in the range 2 -- 200d. Longer periods are harder to explore because they tend to approach the length of each data block ($\sim$200d) and the annual observing cycle. Two strong adjacent peaks emerged in the power spectrum - see Figure~\ref{fig:ogle2}. These peaks are at 4.520d and 4.576d. According to the Corbet diagram for the SMC sources (Corbet et al., 2009) it is very unlikely that SXP214 could have a binary period of this order. In fact, one of the two peaks (4.576d) appears to be a beat period between the other period and the annual sampling of the OGLE III data. This is confirmed by examining just one year of the data set which reveals just the one peak at 4.520d. Furthermore, if this period does not represent a binary period, then the most likely explanation is that it is a Non-Radial Pulsation (NRP) within the Be star (Diago et al., 2008), or, perhaps, the beat of the true NRP period with the daily sampling.

\subsection{Optical spectroscopy}

\begin{figure}
\includegraphics[angle=-0,width=80mm]{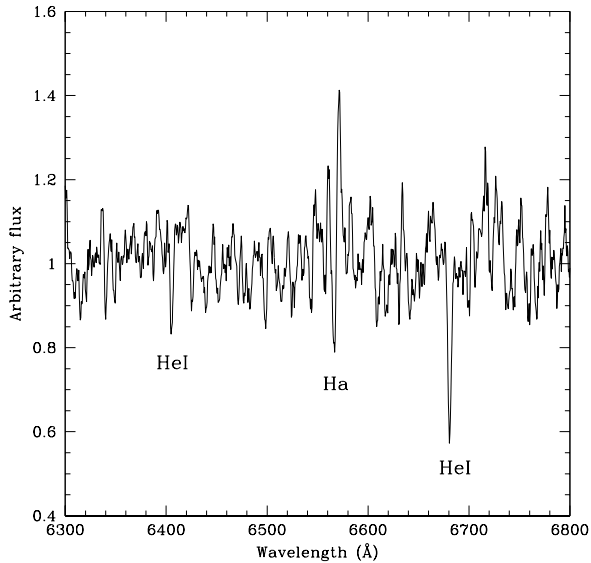}
\caption{Optical spectrum of the region around H$\alpha$. The most obvious line features are indicated. }
\label{fig:ha}
\end{figure}

Spectroscopic observations of the H$\alpha$ region were made on 11
Dec. 2009 (MJD 55176) using the 1.9m telescope of the South
African Astronomical Observatory (SAAO). A 1200 lines per
mm reflection grating blazed at 6800\AA~ was used with the SITe
CCD which is effectively 266×1798 pixels in size, creating a
wavelength coverage of 6200\AA~ to 6900\AA~. The intrinsic resolution
in this mode was 0.42\AA~/pixel. The spectrum resulting from a 2000s integration is shown in
Figure~\ref{fig:ha}.

The signal to noise is poor in this spectrum, but there is definite evidence for a weak feature around H$\alpha$. An H$\alpha$ line emission width of -1.5$\pm$1.0\AA~ is estimated. Higher quality data are definitely required to establish if the line shows any structure. Absorption line features associated with HeI at rest wavelengths of 6406\AA~ and 6676\AA~ are also marked on the figure.

\begin{figure}
\includegraphics[angle=-0,width=80mm]{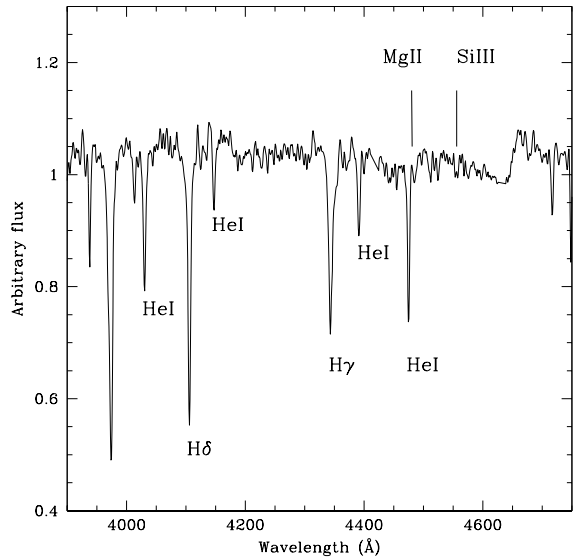}
\caption{Optical spectrum of the spectral classification region.}
\label{fig:blue}
\end{figure}

Additional observations were made using the Integral Field Spectroscopy (IFS) instrument on 28 December, 2010
at Siding Springs Observatory using the 2.3m Advanced Technology Telescope and its Wide
Field Spectrograph (WiFeS). The 1200 second single exposure was made in the central region of
SXP214 at position angle (east of north) 0 degrees under
clear skies with seeing estimated at one arcsec. The exposure was made in classical equal
mode using the RT560 beam splitter and 3000 Volume Phased Holographic (VPH) gratings.
For these gratings, the blue (708 lines mm-1) range includes 3200 -- 5900 \AA.
The data were reduced using the WiFeS data reduction pipeline based on NOAO (National
Optical Astronomy Observatory) IRAF software. This data reduction package was
developed from the Gemini IRAF package (McGregor et al. 1997). Use of the pipeline consists
of four primary tasks: wifes to set environment parameters, wftable to convert
single extension FITS file formats to Multi-Extension FiTS ones and create file lists used
by subsequent steps, wfcal to process calibration frames including bias, flat-field, arc and
wire; and wfreduce to apply calibration files and create data cubes for analysis.
Using QFitsView3\footnote {Written by Thomas Ott and freely available at
www.mpe.mpg.de/~ott/dpuser/index.html}, a spectrum of SXP214 in the blue range was created and is shown in Figure~\ref{fig:blue}.

\section{Discussion and Conclusions}

During the XMM-Newton survey of the SMC we discovered a new hard X-ray transient with a 0.2$-$10 keV luminosity of $7\times10^{35}$erg/s. The precise position derived from the EPIC data allowed us to identify a V=15.3 mag early type star as optical counterpart. An optical spectrum taken at SAAO revealed a weak H$_\alpha$ line which clearly identifies the source as a Be/X-ray binary in the SMC. After SXP11.87, this is the second discovery of such a system during the XMM-Newton SMC survey (Sturm et al 2010).

SXP214 shows X-ray pulsations with a period of 214.05s and a hard (power law) X-ray spectrum (photon index of $\sim$0.65).
Conspicuous are the residuals in the EPIC-pn spectrum at low energies ($\le$1.5keV - see Fig.~\ref{fig:spec}).
This might be caused by instrumental differences at high off-axis angles
or by the contribution of a soft excess (Hickox et al, 2004).
In the latter case, the low count rate and the high absorption do not allow to fit an additional component.
We note that a soft excess originating near the neutron star (accretion column or disk) and absorbed by the same
high column density as the power-law component would require an unrealistic high luminosity of L$\sim$10$^{39}$ erg s$^{-1}$ to cause this signal.
Therefore, if a soft spectral component exists, it more likely originates in an optically thin gas cloud around the neutron star attenuated
by much lower absorption.
Given the relatively low outburst luminosity of SXP214, this is consistent with the conclusions about the origin of soft excesses in spectra of HMXBs drawn by Hickox et al (2004).

At the time of this XMM-Newton detection and other related observations, the source seems to be in an optical low state compared with the past decade. In addition, the three previous XMM-Newton observations of SXP214 have upper limits at least a factor of 10 below the flux level reported here ($7\times10^{35}$erg/s). So at first glance it seems strange that it was finally detected by XMM-Newton when the optical flux was at such a low point. However, placing SXP214 on the Corbet diagram would suggest a binary period of the order 100--200d. It is therefore very probable that the previous observations failed to catch the system with the neutron star at periastron and hence missed any Type I outbursts that may have occurred. The long-term RXTE monitoring (Figure~\ref{fig:xte}) at a good collimator sensitivity clearly shows the lack of any more prolonged ($\ge$1 month) Type II outbursts from this system.

\begin{figure}
\includegraphics[angle=-0,width=80mm]{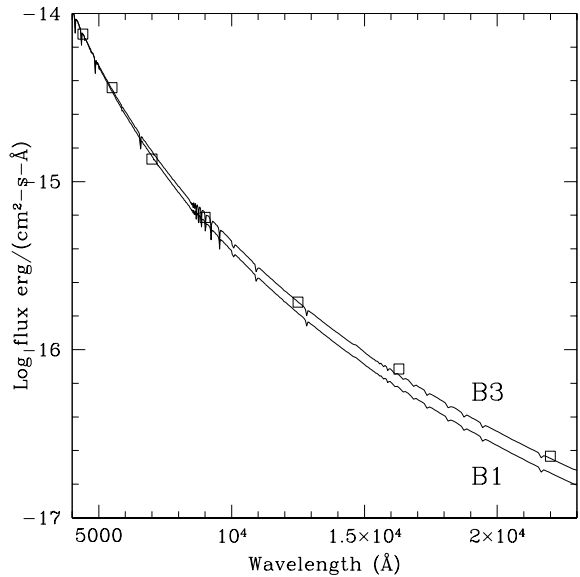}
\caption{Optical and IR photometry (BVRIJHK) de-reddened with E(B-V)=0.09 and compared the Kurucz model atmospheres for B1V and B3V stars. The model atmospheres have been normalised to the B band point.}
\label{fig:flux}
\end{figure}

Not only is the I band flux at a very low value for this system, but the H$\alpha$ equivalent width also implies a very minimal circumstellar disk was present in late 2009.
If Kuruscz model atmospheres (Kuruscz 1979) representing the proposed spectral class range (B1V - B3V) are normalised to the dereddened B band flux (using E(B-V)=0.09 from Schwering \& Israel, 1991), it again becomes clear that there is very little evidence for any substantial IR excess arising from a circumstellar disk - see Figure~\ref{fig:flux}. So anything from B1V with a small disk, to B3V with essentially no disk at all, offers a satisfactory fit to all the optical and IR data.

A more precise spectral classification may be determined from Figure~\ref{fig:blue}. Following the method given in Evans et al (2004) an absence of HeI $\lambda\lambda$4200, 4541 lines defines the class to be later than B0. The weak but approximately equal strength of Mg II ($\lambda$4481) and Si III($\lambda$4553) narrows it down to B2.5, though it could be up to half a class on either side. In the context of the luminosity class, if we take the V band magnitude presented here and correct for the distance modulus (18.9) and the extinction to the SMC (E(B-V)=0.08) an absolute magnitude of -3.83 results. Comparing this to the data in Wegner (1994) predicts a luminosity class III for a B2 star. So the best spectral identification from the data presented here gives B2-B3 III, which adds further support to a minimal circumstellar disk in this system.

In contrast, the column density derived from the X-ray spectrum of
$\sim5 \times 10^{22}$ cm$^{-2}$ is relatively high for Be/X-ray binaries in the SMC (see Fig. 12 in Haberl et al. 2008) and largely exceeds the total SMC HI column density of $1.1 \times 10^{22}$ cm$^{-2}$ (Stanimirovic et al. 1999). This suggests a large amount of source-intrinsic absorption close to the neutron star or that the line of sight to the X-ray source during the XMM-Newton observation passed close to the Be star, implying a large system inclination. An even more extreme case is the Be/X-ray binary pulsar SXP1323 = RX\,J0103.6$-$7201 presented by Eger \& Haberl (2008). During one of more than twenty XMM-Newton observations of  RX\,J0103.6$-$7201 the absorption exceeded $10^{23}$ cm$^{-2}$, completely suppressing the hard power-law component at energies below 2 keV and revealing a strong soft spectral component, likely caused by re-processing in optically thin gas. In the case of SXP214, this soft component is much weaker, consistent with the conclusion from above that only a minimal circumstellar disk was present during the XMM-Newton observation. It is worth noting that within our galaxy columns of up to $10^{24}$ cm$^{-2}$ have been measured for some highly obscured HMXB systems.

\section{Acknowledgements}

R.S. acknowledges support from the BMWI/DLR grant FKZ 50 OR 0907. LJT is supported by a University of Southampton Mayflower Scholarship.
As always, we are grateful to the support staff in SAAO for help in using the 1.9m and IRSF telescopes. The Faulkes Telescope Project is an educational and research arm of the Las Cumbres
Observatory Global Telescope Network (LCOGTN). The OGLE project has received funding from the European Research Council under the European Community's Seventh Framework Programme (FP7/2007-2013)/ERC grant agreement no. 246678 to AU. MDF thanks Australian Government AINSTO AMNRF for grant number 10/11-O-06.

\bsp

\label{lastpage}

\end{document}